\documentclass[preprint]{revtex4}
\usepackage{graphicx}
\usepackage{epstopdf}
\usepackage{amsmath}

\begin{document}
	
\title{Phonon-mediated superconductivity in two-dimensional hydrogenated phosphorus carbide: HPC$_{3}$}
\author{Ya-Ping Li$^1$, Liu Yang$^1$, Hao-Dong Liu$^1$, Na Jiao$^1$, Mei-Yan Ni$^1$, Ning Hao$^{2}$, Hong-Yan Lu$^{1,*}$, and Ping Zhang$^{1,3,*}$}
\affiliation{$^1$School of Physics and Physical Engineering, Qufu Normal University, Qufu 273165, China \\$^2$Anhui Province Key Laboratory of Condensed Matter Physics at Extreme Conditions, High Magnetic Field Laboratory, HFIPS, Chinese Academy of Sciences, Hefei 230031, China \\$^3$Institute of Applied Physics and Computational Mathematics, Beijing 100088, China}
\begin{abstract}  		
	In the recent years, three-dimensional (3D) high-temperature superconductors at ultrahigh pressure have been reported, typical examples are the polyhydrides H$_{3}$S, LaH$_{10}$, and YH$_{9}$, etc. To find high-temperature superconductors in two-dimensional (2D) at atmosphere pressure is another research hotspot. Here, we investigated the possible superconductivity in a hydrogenated monolayer phosphorus carbide based on first-principles calculations. The results reveal that monolayer PC$_{3}$ transforms from a semiconductor to a metal after hydrogenation. Interestingly, the C-$\pi$-bonding band contributes most to the states at the Fermi level. Based on the electron-phonon coupling mechanism, it is found that the electron-phonon coupling constant of HPC$_{3}$ is 0.95, which mainly origins from the coupling of C-$\pi$ electrons with the in-plane vibration modes of C and H. The calculated critical temperature $T_{c}$ is 31.0 K, which is higher than most of the 2D superconductors. By further applying biaxial tensile strain of 3$\%$, the $T_{c}$ can be boosted to 57.3 K, exceeding the McMillan limit. Thus, hydrogenation and strain are effective ways for increasing the superconducting  $T_{c}$ of 2D materials.
	
\end{abstract} 

\maketitle

\section{Introduction}

Since the discovery of superconductor, its irreplaceable superior physical properties and wide range of applications have been the subject of enthusiastic research. Especially, finding superconductors with high critical temperature ($T_{c}$) is the goal that researchers are keen to pursue. In 1968, Ashcroft predicted that the dense metallic hydrogen may be a candidate for high-temperature superconductor \cite{Ashcroft}. The reason is that the minimal mass of H would lead to high Debye frequency, which can increase $T_{c}$ based on the Bardeen-Cooper-Schrieffer (BCS) theory \cite{BCS}. Nevertheless, metallic hydrogen is difficult to synthesize experimentally \cite{Dalladay, Loubeyre, Eremets}. Therefore, researchers consider whether pure hydrogen can be replaced  by compounds with high hydrogen content to explore high-temperature superconductors. Recently, several 3D polyhydrides have been investigated to show high $T_{c}$. For example, hydrogen sulfide H$_{3}$S was theoretically predicted \cite{Duan1} and experimentally confirmed with $T_{c}$ of 203 K at 155 GPa \cite{Drozdov1}. Later, $T_{c}$ was predicted to be 274-286 K for LaH$_{10}$ at 210 GPa, 253-276 K for YH$_{9}$ at 150 GPa, and 305-326 K for YH$_{10}$ at 250 GPa \cite{Liu, Peng}. Among them, high-temperature superconductivity has been observed in LaH$_{10}$ and YH$_{9}$ experimentally \cite{Drozdov2, Somayazulu, Snider1}. More recently, $T_{c}$ of 288 K in a carbonaceous sulfur hydride at 267 GPa was experimentally reported \cite{Snider2}.

However, above 3D high-temperature superconductors and can only be realized at high pressure, which greatly limits the application of the superconductors. On the other hand, 2D superconductors may have good applications in constructing nano superconducting devices. Thus, exploring 2D high-temperature superconductors at atmosphere pressure attracts a lot of attention theses years. Till now, some progress has been made in the research on the superconductivity in 2D materials. For example, lithium and calcium deposited graphene LiC$_{6}$ and CaC$_{6}$ were predicted to be phonon-mediated superconductors with $T_{c}$ of 8.1 K and 1.4 K, respectively \cite{Profeta}. Later,  LiC$_{6}$ was experimentally proved to be a superconductor with $T_{c}$ of 5.9 K \cite{Ludbrook}. Moreover, it was predicted that doping and applying biaxial tensile strain can increase the $T_{c}$ of graphene to 31.6 K \cite{Duan}. The magic angle bilayer graphene has been found to be a superconductor with $T_{c}$ of 1.7 K \cite{CaoY1, CaoY2}. The aluminum-deposited graphene (AlC$_{8}$) was also predicted to be a superconductor with $T_{c}$ of 22.2 K by doping and stretching \cite{Lu}. Besides graphene, there are other 2D superconductors which will be show in Table I.  However, the $T_{c}$ of these superconductors are relatively low. Whether 2D high-temperature superconductivity can be achieved through other methods is a hotspot of the current research.

Hydrogenation, which can modify the electronic properties of materials, has attracted more and more attention in the recent years. Typically, hydrogenation can modify the electronic properties of graphene \cite{sofo, Graphane, luhongyan1, luhongyan2, luhongyan3, Sahin}. For instance, fully hydrogenated graphene, called graphane, has been theoretically predicted \cite{sofo} and experimentally synthesized \cite{Graphane} to be an insulator with a direct band gap of 3.5 eV.  Ferromagnetism and superconductivity with possible $p + i p$ pairing symmetry was predicted in partially hydrogenated graphene \cite{luhongyan2}. In addition, theoretical calculations show that doped graphane is a high-temperature electron-phonon superconductor with $T_{c}$ above 90 K \cite{Savini}. Similarly, it was predicted that hydrogenated monolayer magnesium diboride ${(}$MgB$_{2}$${)}$ can be a superconductor with $T_{c}$ of 67 K \cite{Bekaert}. Thus, it is of great significance to study the effect of hydrogenation on the electronic properties and possible high-temperature superconductivity in 2D materials at atmosphere pressure.

In the recent years, 2D monolayer phosphorus carbides have been investigated to show interesting physical properties. For example, several kinds of phosphorus carbides, $\alpha$$_{0}$-PC, $\alpha$$_{1}$-PC, $\alpha$$_{2}$-PC, $\beta$$_{0}$-PC, $\beta$$_{1}$-PC, and $\beta$$_{2}$-PC,  were predicted to be in an atomically thin layer and to be metal, semimetal, or semiconductors, respectively \cite{Guan}. $\beta$$_{0}$-PC has also been predicted to be a superconductor with $T_{c}$ of 13 K \cite{wang}. Monolayer PC$_{6}$ was predicted to be a semiconductor with a direct bandgap of 0.84 eV and anisotropic carrier mobility \cite{Yu}. Monolayer PC$_{3}$, a semiconductor with an indirect bandgap of 1.46 eV, was proposed to be a promising thermoelectric material \cite{Rajput} or to be used as K ion battery anode \cite{Song}. Among the phosphorus carbides, monolayer PC$_{3}$ shows similar lattice structure as graphene and has favorable advantages of high symmetry and light mass. Therefore, it is anticipated that hydrogenated monolayer PC$_{3}$ may also show superconductivity. Based on first-principles calculations, it was predicted that monolayer pristine HPC$_{3}$ is a superconductor with $T_{c}$ of 31.0 K. By further applying biaxial tensile strainand, the $T_{c}$ can be boosted to 57.3 K, exceeding the McMillan limit. 

The rest of this article is organized as follows. In Sec. II, we describe the computational details. In Sec. III, we present the results and discussions. Firstly, we show the stability of HPC$_{3}$. Second, we discuss the electronic structure, including the band structure, Density of states (DOS) and Fermi surface (FS) of HPC$_{3}$. Third, the phonon and electron-phonon coupling of HPC$_{3}$ are calculated and superconducting $T_{c}$ is further calculated. Sec. IV is the conclusion of the work.

\section{Results and discussions}
\subsection{Lattice structure and stability}

\begin{figure}
	\centering
	\label{fig:stucture}
	\includegraphics[width=16cm, height=7cm]{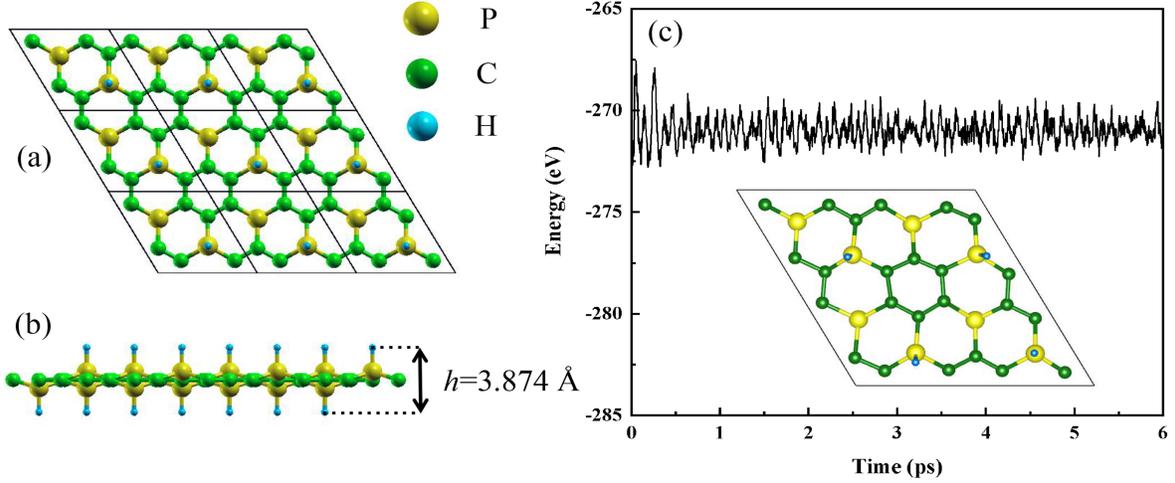}
	\caption{(a) The top view of HPC$_{3}$. (b) The side view of the HPC$_{3}$. (c) The variation of the free energy in the AIMD simulations during the time scale of 6 ps along with the last frame of photographs at 700 K. The yellow, green, and blue spheres represent phosphorus, carbon, and hydrogen atoms, respectively. The solid black line represent the unit cell.}
\end{figure}

The crystal structure of HPC$_{3}$ is shown in Fig. 1(a). The yellow, green, and blue spheres represent carbon, phosphorus, and hydrogen atoms, respectively. It is just like a honeycomb ring of C atoms surrounded by six P atoms on the outside. In the unit cell, there are two P atoms, six C atoms and two H atoms, with each H atom bonding with a P atom along different sides of the C plane, which can be easily seen from the side view of the HPC$_{3}$ in Fig. 1(b). The lattice constant is optimized to 5.408 $\textmd{\AA}$. From the side view, it can be clearly seen that the lattice structure is wrinkled, and the height between hydrogen atoms on different sides ($h$) is 3.874 $\textmd{\AA}$, and P-H, P-C, and C-C length are 1.445 $\textmd{\AA}$, 1.765 $\textmd{\AA}$, 1.428 $\textmd{\AA}$, respectively. In all the calculations, a vacuum space of 20 $\textmd{\AA}$ was used to avoid the interactions between adjacent layers. In PC$_{3}$, the C atoms are in the $sp^{2}$ configuration as in graphene, and P atoms adopt a $sp^{3}$ hybridization, in which three hybrid orbitals form covalent bonds with the neighboring C atoms and one hybrid orbital is filled with a lone electron pair. After hybridization, one electron of the lone electron pair forms covalent bond with H atom, leaving the other one a $\pi$-like electron.   

For the stability of HPC$_{3}$, it was studied from two aspects. One is the thermodynamic stability, and the other is the dynamical stability. Regarding thermodynamic stability, it was proved from the ab-initio molecular dynamics (AIMD) simulations. A 2 $\times$ 2 $\times$ 1 supercell was used to minimize the effect of periodic boundary condition. The variation of free energy in the AIMD simulations within 6 ps and the last frame of the photographs are shown in Fig. 1(c). The results show that the integrity of the structure has not changed even at 700 K, proving its thermodynamic stability. Moreover, the phonon spectrum of HPC$_{3}$ shows no imaginary frequency (it will be shown later), indicating that it is dynamical stable. Thus, the structure of HPC$_{3}$ is stable at room temperature, satisfying both thermodynamic and dynamical stability.

\subsection{Electronic structure}

\begin{figure}
	\centering
	\label{fig:dos}
	\includegraphics[width=11.5cm, height=10cm]{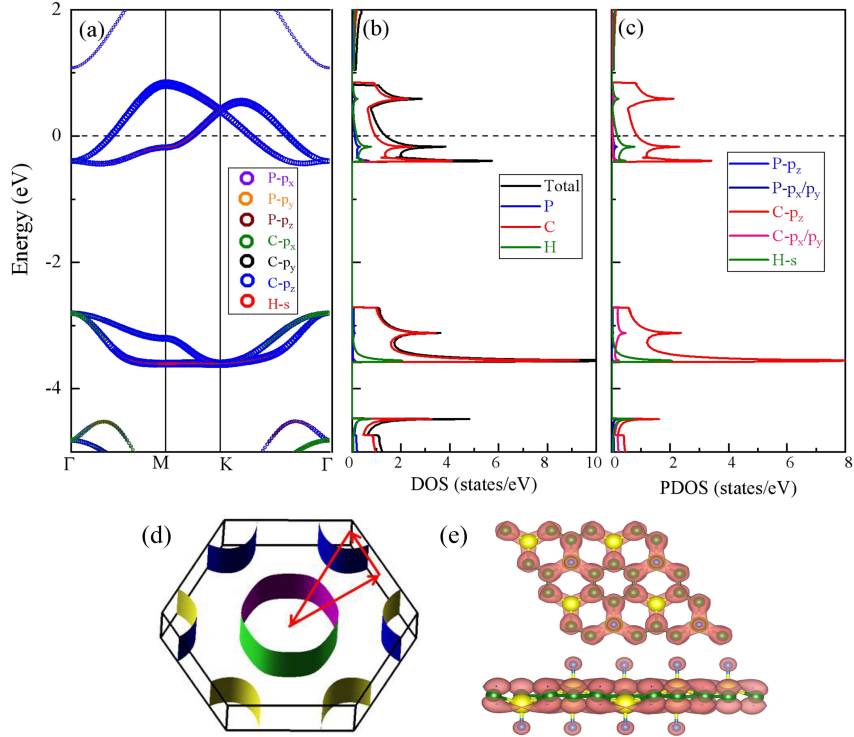}
	\caption{(a) Orbital-projected electronic band structure of HPC$_{3}$ along high-symmetry line $\Gamma$-M-K-$\Gamma$. (b) The total DOS of HPC$_{3}$ and the total DOS of P, C and H atoms. (c) The partial density of states (PDOS) of HPC$_{3}$. (d) The side view of FS of HPC$_{3}$. (e) Charge density of HPC$_{3}$ in the range of -1 to 1 eV near the Fermi level. The Fermi level in (a) (b) and (c) are set to zero.}
	
\end{figure}
\begin{figure*}
	\centering
	\label{fig:phonon}
	\includegraphics[width=16 cm, height=7cm]{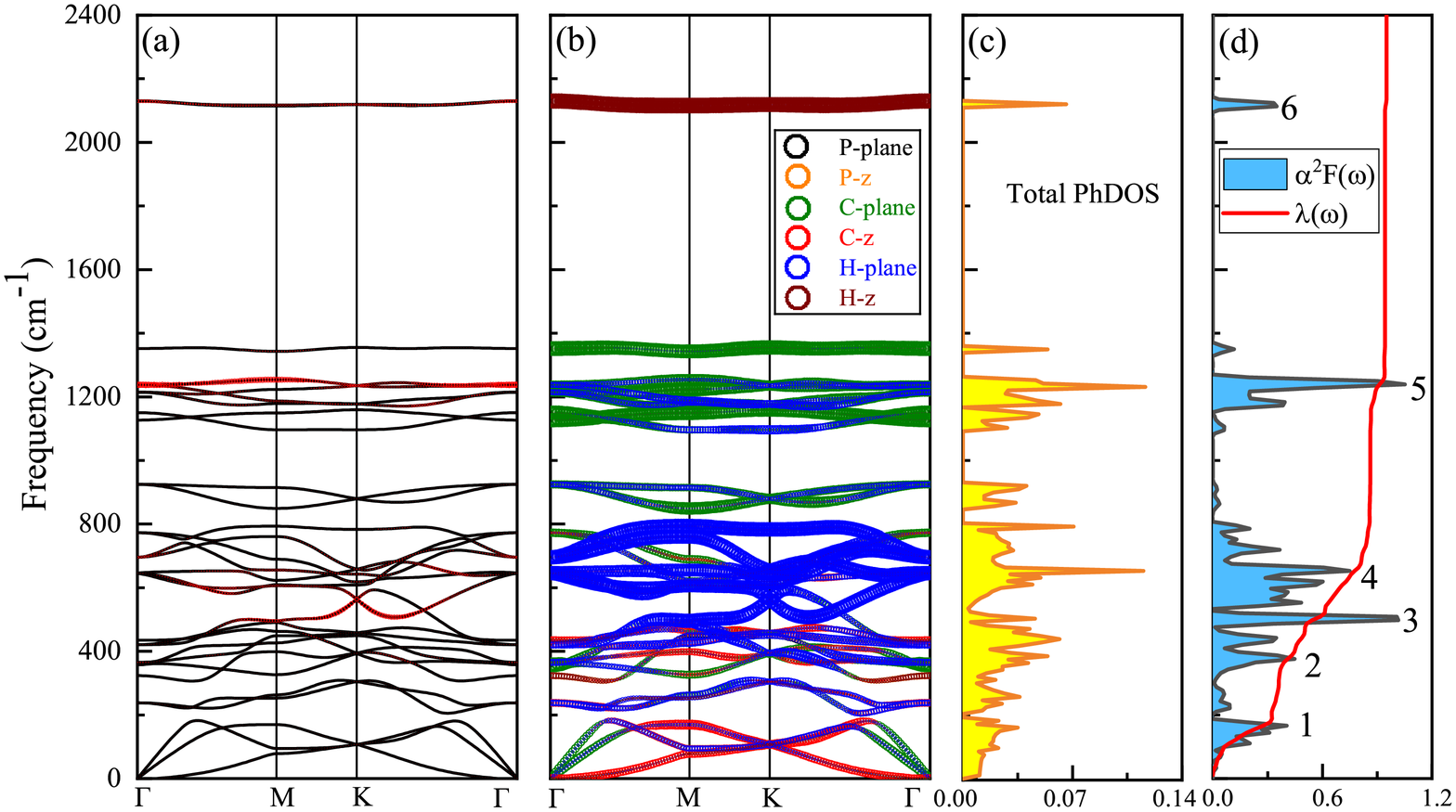}
	\caption{(a) Phonon dispersion of HPC$_{3}$. The red hollow circles indicate the phonon linewidth $\gamma_{\textbf{q}\nu}$. (b) Phonon dispersion of HPC$_{3}$ weighted by the vibration modes of P, C and H atoms. The black, orange, green, red, blue, and wine red hollow circles indicate P horizontal, P vertical, C horizontal, C vertical, H horizontal, and H vertical modes, respectively. (c) Total PhDOS and (d) Eliashberg spectral function $\alpha^{2}F (\omega)$ and cumulative frequency-dependent EPC function $\lambda (\omega)$.}
\end{figure*}

We firstly calculated the electronic band structure and DOS of the monolayer PC$_{3}$ based on first-principles calculations, and found it is an indirect bandgap semiconductor with the gap of 1.488 eV (Fig. S1 \cite{supplement}), consistent with the former result \cite{Rajput}. For the electronic structure of HPC$_{3}$, the band structure, DOS, partial density of states (PDOS) and FS were studied, with the results shown in Fig. 2. From the band structure in Fig. 2(a), it is shown that two bands cross the Fermi level, proving that HPC$_{3}$ is a metal. Fig. 2(b) is the total DOS and the total DOS of each element. It can be seen that the states at the Fermi level are mainly contributed by C atoms, the contribution of H atoms is less, and that of P atoms is the smallest. The PDOS of HPC$_{3}$ is shown in Fig. 2(c), which confirms that the $p_{z}$ orbital of C atoms contributes the most near the Fermi level, and then is the $1s$ orbital of H. Fig. 2(d) is the Fermi surface of HPC$_{3}$, where an electron pocket around $\Gamma$ and six hole pockets aroud $K$ can be clearly seen. Furthermore, the charge density in the range of -1 to 1 eV near the Fermi level was plotted, which is shown in Fig. 2(e). It is seen that the charge mainly origins from the $p_{z}$ orbital of the C atoms, i.e., the $\pi$-bonding electrons. It was previously proposed that the superconductivity mainly arises from the coupling of $\sigma$-bonding bands with phonon \cite{GaoMiao1,GaoMiao2}. For HPC$_{3}$, since the two $\pi$-bonding bands have significant impacts on the electronic DOS at the Fermi level, we propose that the existence of the $\pi$-bonding bands may also result in strong electron-phonon coupling, leading to superconductivity with high $T_{c}$. 

\subsection{Electron-phonon coupling and possible superconductivity}

Since HPC$_{3}$ is a metal, we study its possible phonon-mediated superconductivity. We firstly study the phonon properties of HPC$_{3}$. Fig. 3(a) shows the phonon dispersion with phonon linewidth $\gamma_{\textbf{q}\nu}$ (red hollow circles). There are ten atoms in the unit cell, leading to thirty phonon bands. It includes three acoustic phonon modes and twenty-seven optical phonon modes. It shows a wide range of frequency extending up to about 2130 cm$^{-1}$, and no imaginary frequency appears in the low energy, justifying its dynamical stability. The vibration modes at $\Gamma$ point are shown in Fig. S2. From the decomposition of the phonon spectrum with respect to P, C, and H atomic vibrations, as indicated in Fig. 3(b), it is seen that the main contribution to the acoustic branches below 200 cm$^{-1}$ are the out-of-plane vibration of C atoms and the in-plane vibration of H atoms. From 200 to 472 cm$^{-1}$, the mainly contribution is from the out-of-plane vibration of C atoms, the in-plane vibration of H atoms, and a small amount of the in-plane vibration of P atoms. In the range of 472 $<$ $\omega$ $<$ 800 cm$^{-1}$, the main contribution is from the in-plane vibration of H atoms and a small amount of the in-plane vibration of C atoms. From 800 to 1347 cm$^{-1}$, the main contribution is from the in-plane vibration of C atoms and a small amount of the in-plane vibration of H atoms. The out-of-plane vibration of H atoms occupies the high frequencies around 2130 cm$^{-1}$. 

\begin{table*}
	\renewcommand\arraystretch{0.8}
	\centering
	\caption{\label{tab:table2} List of superconducting parameters required for the prediction of $T_{c}$ for some reported 2D phonon-mediated superconductors. This table includes data of doping, biaxial tensile strain $\varepsilon$, Coulomb pseudopotential $\mu^{*}$, logarithmic averaged phonon frequency $\omega$$_{log}$ (in K), total EPC constant $\lambda$, and estimated $T_{c}$ (in K). Experimental $T_{c}$ values are also noted in the table.}
	~\\ 
	\scalebox{0.8}{
		\begin{tabular}{cccccccc}
			\hline
			\hline
			2D Materials      & doping          & $\varepsilon$  & $\mu^{*}$      & $\omega$$_{log}$ (K)    & $\lambda$   & $T_{c}$ (K)  & Ref. \\
			\hline 
			graphene         &4.65*10$^{14}$ cm$^{-2}$& 16.5$\%$&  0.10          &  289.06                 &  1.45       &30.2         & \cite{Duan} \\
			graphane	     &     10$\%$ $p$-doping  &     0.0 &  0.13          &  909.35                 &  1.45       &91.1         & \cite{Savini}\\           
			stanene          &  Li deposition  &     0.0        &   0.13         &   63.60                 & 0.65        & 1.4         & \cite{st}  \\
			silicene         & 3.51*10$^{14}$ cm$^{-2}$ & 5$\%$ &   0.10         &   304.25                & 1.08        & 16.4        & \cite{si} \\
			phosphorene      & 1.1 $e$/cell    &     0.0        &   0.10         &    -                    & 1.20        & 11.2        & \cite{p}  \\	
			$\beta$$_{12}$ borophene   &      0.0        &     0.0        &   0.10         &   323.29                & 0.89        & 18.7        & \cite{b1}  \\ 
			$\chi$$_{3}$ borophene    &      0.0        &     0.0        &   0.10         &   383.96                & 0.95        & 24.7        & \cite{b1}  \\
			B$_{2}$C         &      0.0        &     0.0        &   0.10         &   314.80                & 0.92        & 19.2        & \cite{b2c}  \\
			CaC$_{6}$        &      0.0        &     0.0        &   0.115        &   445.64                & 0.40        & 1.4         & \cite{Profeta}  \\
			LiC$_{6}$        &      0.0        &     0.0        &  0.115         &   399.48                & 0.61        & 8.1         & \cite{Profeta}  \\
			LiC$_{6}$        &      0.0        &     0.0        &     -          &     -                   & 0.58 $\pm$ 0.05 & 5.9 [Exp.]  & \cite{Ludbrook}  \\
			AlC$_{8}$        &  0.20 $e$/cell  &     12$\%$     &  0.10          &   188.452               & 1.557       & 22.23       & \cite{Lu}  \\
			monolayer 2H-NbSe$_{2}$    &      0.0        &     0.0        &    -           &     -                   & 0.75        & 3.1 [Exp.]  & \cite{nb}  \\
			monolayer 2H-NbSe$_{2}$    &      0.0        &     0.0        &  0.16/0.15     &   134.3/144.5           & 0.84/0.67   & 4.5/2.7     & \cite{nb1}, \cite{nb2}  \\
			Mo$_{2}$C        &      0.0        &     0.0        &   0.10         &    -                    & 0.63        &5.9          &\cite{J.J}\\
			B$_{2}$O         &      0.0        &     0.0        &   0.10         &   250.0                 & 0.75        & 10.35       & \cite{b2o}  \\
		bilayer	MoS$_{2}$        &Na-intercalation&     7$\%$       &   0.10         &   135                   & 1.05        & 10.05      & \cite{Shuai Dong1}  \\
		bilayer	blue phosphorus  &Li-intercalation&    0.0         &   0.10         &   221.3                 & 1.2         & 20.4        & \cite{Shuai Dong2}  \\
		bilayer	$\beta$-Sb   &Ca-intercalation &    0.0         &   0.10         &    -                    & 0.89        & 7.2         & \cite{Shuai Dong3}  \\
			monolayer MgB$_{2}$  &      0.0        &     0.0        &   0.13         &    -                    & 0.68        & 20          & \cite{mgb2}  \\	
			hydrogenated monolayer MgB$_{2}$&   0.0        &     0.0        &   0.13         &    -                    & 1.46        & 67          & \cite{Bekaert}\\
			$\beta$$_{0}$-PC         &      0.0        &     0.0        &   0.10         &  118.0                  & 1.48        & 13.35       & \cite{wang}  \\
			&      0.0        &     0.0        &   0.10         &  482.84                 & 0.95        &31.0         &   Our work   \\
			HPC$_{3}$      &      0.0        &    1$\%$       &   0.10         &  497.73                 & 1.05        &37.3         &   Our work    \\
			&      0.0        &    2$\%$       &   0.10         &  475.57                 & 1.24        &44.4         &   Our work     \\
			&      0.0        &    3$\%$       &   0.10         &  402.84                 & 1.65        &57.3         &   Our work      \\
			\hline 
			\hline                             
		\end{tabular}
	}
\end{table*}

The total phonon density of state (PhDOS) of HPC$_{3}$ is shown in Fig. 3(c). The Eliashberg spectral function $\alpha^{2}F (\omega)$ and cumulative frequency-dependent EPC function $\lambda (\omega)$ are displayed in Fig. 3(d). By comparing the total PhDOS in Fig. 3(c) and the Eliashberg spectral function $\alpha^{2}F (\omega)$ in Fig. 3(d), it is seen that the peaks of them are basically at the same frequency. Besides, the significant phonon linewidths around 472 and 1272 cm$^{-1}$ lead to the 3rd and 5th peaks of Eliashberg spectral function, respectively. This is understandable from the definition of $\alpha^{2}F (\omega)$ (Eq. 1). As is shown in Fig. 3(d), the total EPC constant $\lambda$ is 0.95, which originates from the coupling between phonons and the $\pi$ electrons of C atoms and a small part of $s$ electron of H atoms. At low frequency 0 $<$ $\omega$ $<$ 472 cm$^{-1}$, the 1st and 2nd peaks of $\alpha^{2}F (\omega)$ are responsible for the gradual enhancement of $\lambda (\omega)$ to 0.45, accounting for 47$\%$ of the total EPC. In the range of 472 $<$ $\omega$ $<$ 800 cm$^{-1}$, the 3rd and 4th peaks are responsible for the $\lambda$ strength 0.4, accounting for 42$\%$ of the total EPC. Around 1272 cm$^{-1}$, the 5th peak is responsible for the $\lambda$ strength 0.09, accounting for 10$\%$ of the total EPC. At high frequency around 2130 cm$^{-1}$, the 6th peak of $\alpha^{2}F (\omega)$ contributes only 0.01 (1$\%$) of the total EPC. Moreover, based on the Eliashberg spectral function $\alpha^{2}F (\omega)$, the calculated logarithmically averaged phonon frequency $\omega_{log}$ is 482.84 K. Using the Allen-Dynes modified McMillan equation \cite{McMillan}, the $T_{c}$ of HPC$_{3}$ is 31.0 K, which is higher than most of the 2D superconductors, except for the doped graphane \cite{Savini} and the hydrogenated monolayer MgB$_{2}$ \cite{Bekaert}, with the results shown in Table I.

\begin{figure}
	\centering
	\label{fig:phdos}
	\includegraphics[width=10cm, height=8cm]{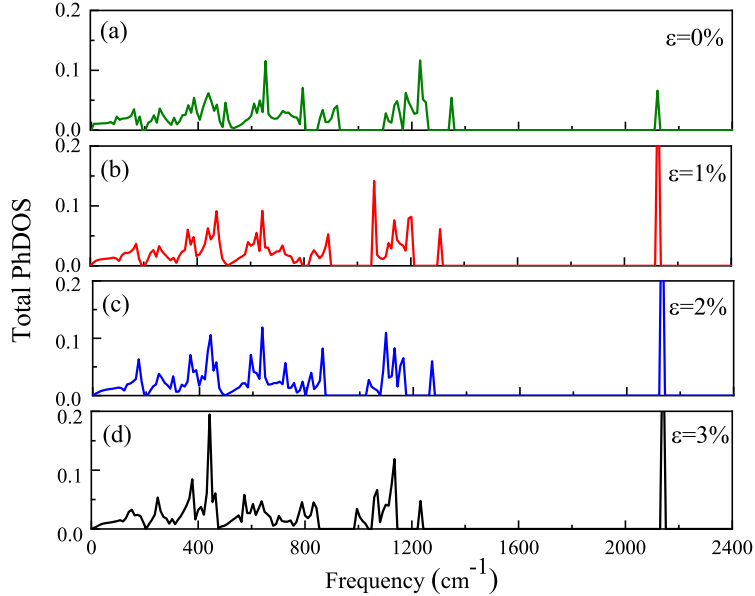}
	\caption{Total PhDOS for (a) pristine, (b-d) 1$\%$, 2$\%$, 3$\%$ biaxial tensile strained HPC$_{3}$.}
\end{figure}

\begin{figure}
	\centering
	\label{fig:Elishberg}
	\includegraphics[width=10cm, height=8cm]{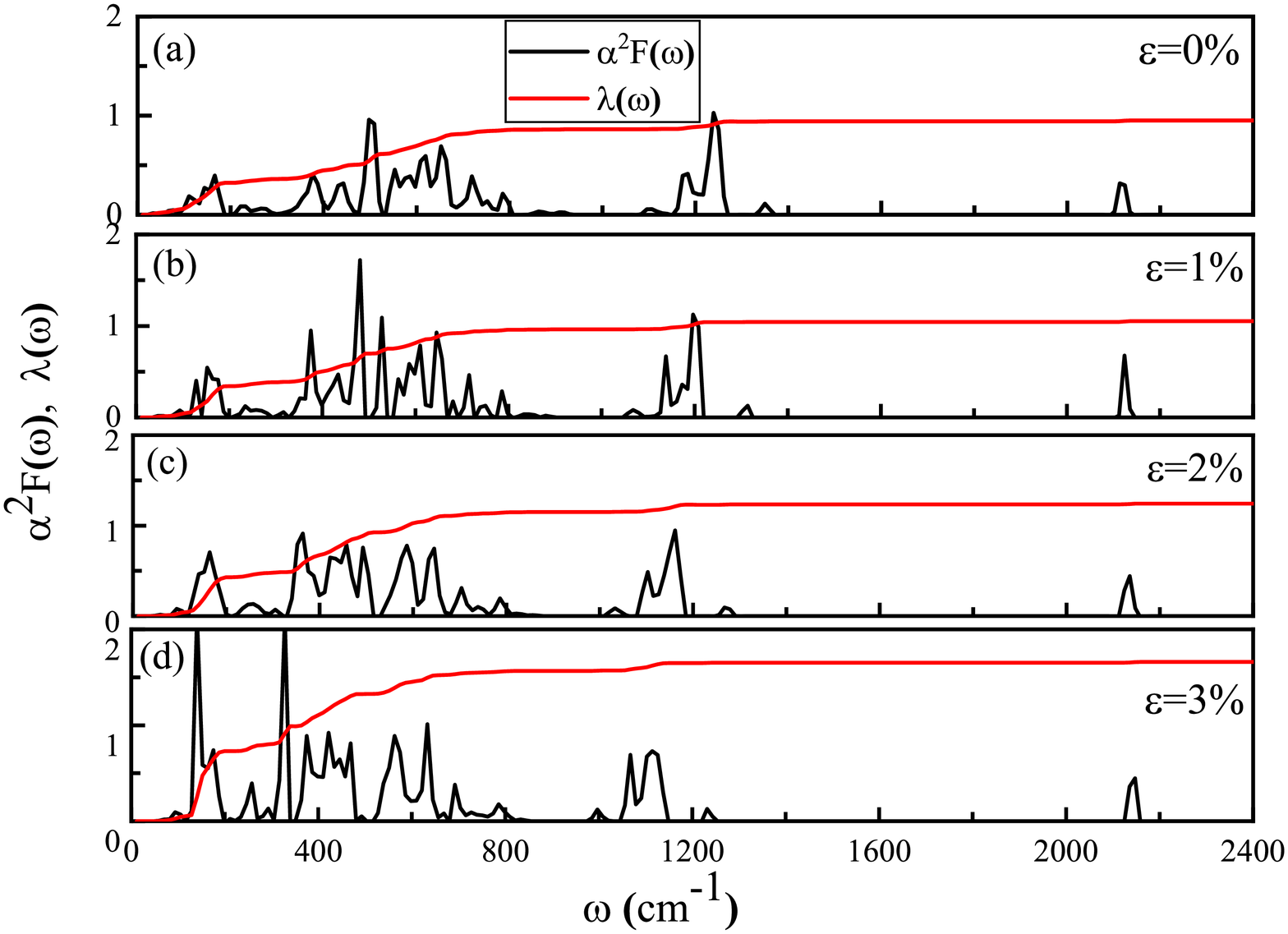}
	\caption{Eliashberg spectral function $\alpha^{2}F (\omega)$ and EPC function $\lambda (\omega)$ for (a) pristine, (b-d) 1$\%$, 2$\%$, 3$\%$ biaxial tensile strained HPC$_{3}$.}
\end{figure}

Furthermore, we considered whether the application of biaxial tensile strain will increase the $T_{c}$ of HPC$_{3}$, which are the cases in doped or metal-deposited graphene \cite{Duan, Lu} and doped monolayer hexagonal boron nitride \cite{GaoMiao3}. The tensile strain was applied along the two basis vector directions, and the relative increase of lattice constant was defined as $\varepsilon=(a-a_{0})/a_{0}$, where $a_{0}$ and $a$ are the lattice constant of pristine and strained HPC$_{3}$, respectively. By applying a serial of biaxial tensile strains, it is found that for strains greater than 4$\%$, there is imaginary frequency in the phonon spectra, implying the lattice is no longer stable. Thus, we show the effect of tensile strain on phonon and superconductivity at $\varepsilon$ = 1$\%$, 2$\%$, and 3$\%$.

Fig. 4 shows the total PhDOS for pristine and biaxial tensile strained HPC$_{3}$. With the increase of tensile strain, there is obvious phonon softening for low and intermediate frequencies, whereas the high-frequency part barely changes. This greatly enhances the electron-phonon coupling, similar behavior has been found in other 2D superconductors \cite{Duan, Lu, GaoMiao3}. Fig. 5 shows the Eliashberg spectral function $\alpha^{2}F (\omega)$ and EPC function $\lambda (\omega)$ for pristine and biaxial tensile strained HPC$_{3}$. The calculated $\omega$$_{log}$, $\lambda$, and $T_{c}$ for the pristine and tensile strained cases are listed in Table I. It is seen that with the increase of tensile strain, $\omega$$_{log}$ decreases but $\lambda$ increases significantly, leading to the increase of $T_{c}$. It is worth mentioning that at $\varepsilon$ = 2$\%$, the $T_{c}$ is 44.4 K, exceeding the McMillan limit. At $\varepsilon$ = 3$\%$, the total EPC $\lambda$ (1.65) is greater than 1.5. Thus, it is necessary to consider the correction factors. By solving the Eq. (7) to Eq. (10), the obtained correction factors are $f_{1}$=1.102 and $f_{2}$=1.037, and the corrected $T_{c}$ is 57.25 K. It is known that the tensile strain up to 25$\%$ can be achieved for 2D materials\cite{lee}. Thus, the predicted monolayer HPC$_{3}$ and its strained cases are highly possible to be realized in future experiments. More experimental and theoretical researches on 2D HPC$_{3}$ are expected.

\section{Conclusion}

In summary, we have calculated the electronic structure, electron-phonon coupling and possible superconductivity of HPC$_{3}$ by using the first-principles calculation. It is found that hydrogenation changes monolayer PC$_{3}$ from a semiconductor to a metal and shows phonon-mediated superconductivity. The superconductivity of HPC$_{3}$ mainly originates from the strong coupling between the $\pi$ electrons of C atoms and the in-plane vibration modes of C and H atoms. The transition temperature of HPC$_{3}$ is 31.0 K, which is higher than most of other 2D superconductors. Under the biaxial tensile strain, the $T_{c}$ can be boosted to 57.3 K, exceeding the McMillan limit. It is anticipated that the predicted monolayer HPC$_{3}$ and its strained cases can be realized in future experiments. 

\section{METHODS}

All calculations related to HPC$_{3}$ in this article were performed in the framework of density functional theory (DFT), as implemented in the Vienna ab-initio simulation package (VASP) \cite{vasp} and the Quantum Espresso (QE) program \cite{QE}. The interaction between electrons and ions was realized by the Projector-Augmented-Wave method (PAW) \cite{PAW}, and the exchange-correlation potentials was treated using the generalized gradient approximation (GGA) with Perdew-Burke-Ernzerhof parametrization (PBE) \cite{PBE}. Both the lattice parameters and the atom positions were relaxed to obtained the optimized structure. The cutoffs for wave functions and charge density were set as 80 Ry and 800 Ry, respectively. Electronic integration was carried out on a 24 $\times$ 24 $\times$ 1 $k$-point grid. For the calculation of the FS and DOS, the $k$ points of 60 $\times$ 60 $\times$ 5 and 48 $\times$ 48 $\times$ 1 were used for calculation, respectively. The phonon and electron-phonon coupling were calculated on a 12 $\times$ 12 $\times$ 1 $q$-point grid, and a denser 48 $\times$ 48 $\times$ 1 $k$-point grid was used for evaluating an accurate electron-phonon interaction matrix.

The total electron-phonon coupling (EPC) constant $\lambda$ can be obtained via isotropic Eliashberg spectral function \cite{ref:41,ref:42,McMillan}
\begin{eqnarray}
	\alpha^{2}F(\omega)=\frac{1}{{2\pi}{N(E_{F})}}\sum_{\mathbf{q}\nu}\delta(\omega-\omega_{\mathbf{q}\nu})\frac{\gamma_{\mathbf{q}\nu}}{\hbar\omega_{\mathbf{q}\nu}},
\end{eqnarray}
\begin{eqnarray}
	\lambda=2\int_{0}^{\infty}\frac{\alpha^{2}F(\omega)}{\omega}\,d\omega=\sum_{\mathbf{q}\nu}^{}\lambda_{\mathbf{q}\nu},
\end{eqnarray}
where $\alpha^{2}F$($\omega$) is Eliashberg spectral function and N($E_{F}$) is the DOS at the Fermi level, $\omega_{\mathbf{q}\nu}$ is the phonon frequency of the $\nu$th phonon mode with wave vector $\mathbf{q}$, and $\gamma_{\mathbf{q}\nu}$ is the phonon linewidth \cite{ref:41,ref:42,McMillan}. The $\gamma_{\mathbf{q}\nu}$ can be estimated by
\begin{eqnarray}
	\begin{split}
		\gamma_{\mathbf{q}\nu}=\frac{2\pi\omega_{\mathbf{q}\nu}}{\Omega_{BZ}}\ \sum_{\mathbf{k},n,m}\lvert g^{\nu}_{\mathbf{k}n,\mathbf{k}+\mathbf{q}m}\rvert^{2}\delta(\epsilon_{\mathbf{k}n}-E_{F})\\\delta(\epsilon_{\mathbf{k}+\mathbf{q}m}-E_{F}),
	\end{split}
\end{eqnarray}
where $\Omega$$_{BZ}$ is the volume of the BZ, $\epsilon_{\mathbf{k}n}$ and $\epsilon_{\mathbf{k}+\mathbf{q}m}$ indicate the Kohn-Sham energy, and $g^{\nu}_{\mathbf{k}n,\mathbf{k}+\mathbf{q}m}$ represents the screened electron-phonon matrix element. $\lambda_{\mathbf{q}\nu}$ is the EPC constant for phonon mode $\mathbf{q}\nu$, which is defined as
\begin{eqnarray}
	\lambda_{\mathbf{q}\nu}=\frac{\gamma_{\mathbf{q}\nu}}{\pi\hbar N(E_{F})\omega^{2}_{\mathbf{q}\nu}}.
\end{eqnarray}
$T_{c}$ is estimated by the Allen-Dynes modified McMillan equation \cite{McMillan}
\begin{eqnarray}
	T_{c}=\frac{\omega_{log}}{1.2}exp[\frac{-1.04(1+\lambda)}{\lambda-\mu^{*}(1+0.62\lambda)}].
\end{eqnarray}
The hysteretic Coulomb pseudopotential $\mu^{*}$ in Eq. (4) is set to 0.1 and the logarithmic average of the phonon frequencies $\omega_{log}$ is defined as
\begin{eqnarray}
	\omega_{log}=exp[\frac{2}{\lambda}\int_{0}^{\omega_{max}}\alpha^{2}F(\omega)\frac{log\;\omega}{\omega}d\omega].
\end{eqnarray}
For the strong EPC cases, i.e., $\lambda$ $>$1.5, $T_{c}$ is estimated by \cite{McMillan}
\begin{eqnarray}
	T_{c}=f_{1}f_{2}\frac{\omega_{log}}{1.2}exp[\frac{-1.04(1+\lambda)}{\lambda-\mu^{*}(1+0.62\lambda)}].
\end{eqnarray}
Here, $f_{1}$ and $f_{2}$ are strong-coupling correction factor and shape correction factor, respectively, with
\begin{eqnarray}
	f_{1}=\{1+[\frac{\lambda}{2.46(1+3.8\mu^{*})}]^{3/2}\}^{1/3},
\end{eqnarray}	
\begin{eqnarray}
	f_{2}=1+\frac{[(\omega_{2}/\omega_{log})-1]\lambda^{2}}{\lambda^{2}+3.312(1+6.3\mu^{*})^{2}(\omega_{2}/\omega_{log})^{2}},
\end{eqnarray}
in which $\omega_{2}$ is defined as	
\begin{eqnarray}
	\omega_{2}=[\frac{2}{\lambda}\int_{0}^{\omega_{max}}\alpha^{2}F(\omega){\omega}d\omega]^{1/2}. 
\end{eqnarray}

\section{DATA AVAILABILITY}

The numerical datasets used in the analysis in this study, and in the figures of this work, are available from the corresponding author on reasonable request.

\section{CODE AVAILABILITY}

The codes that were used here are available upon request to the corresponding author.

\vspace{0.5cm} Corresponding Authors $^*$E-mail: hylu@qfnu.edu.cn, pzhang2012@qq.com

\begin{acknowledgements}
	Y.P.L. thanks helpful discussion with Fa-Wei Zheng. This work is supported by the National Natural Science Foundation of China (Grant  Nos. 12074213, 11574108, and 12104253), the Major Basic Program of Natural Science Foundation of Shandong Province (Grant No. ZR2021ZD01), and the Project of Introduction and Cultivation for Young Innovative Talents in Colleges and Universities of Shandong Province. N.H. is supported by the National Natural Science Foundation of China (Grant  Nos. 12022413, and 11674331), the “Strategic Priority Research Program (B)” of the Chinese Academy of Sciences (Grant No. XDB33030100), the ‘100 Talents Project’of the Chinese Academy of Sciences, the Collaborative Innovation Program of Hefei Science Center, CAS (Grant No. 2020HSC-CIP002), and the CASHIPS Director's Fund (Grant No. BJPY2019B03).
\end{acknowledgements}

\section{AUTHOR CONTRIBUTIONS}

 H. Y. Lu and P. Zhang conceived the project. Y. P. Li, L. Yang, and H. D. Liu did the calculations. N. Jiao, M. Y. Ni, and N. Hao conducted the data analysis. All authors contributed to the discussion and writing of the paper.

\section{COMPETING INTERESTS}

The authors declare no competing interests.

\end{document}